\documentstyle[12pt]{article}
\def\lapproxeq{\lower .7ex\hbox{$\;\stackrel{\textstyle
<}{\sim}\;$}} \def\gapproxeq{\lower
.7ex\hbox{$\;\stackrel{\textstyle >}{\sim}\;$}}

\begin{document}

\titlepage

\begin{flushright}  RAL-93-092 \\ November 1993
\end{flushright}

\begin{center}
\vspace*{2cm}
{\large \bf Improved QCD sum rule estimates of the higher twist
contributions to polarised and unpolarised nucleon structure
functions}
\end{center}

\vspace*{.75cm}
\begin{center}
Graham G.\ Ross\footnote {SERC Advanced Fellow} \\
Department of Theoretical Physics, University of Oxford, \\
Oxford, UK.  \\ \end{center}

\begin{center}
and
\end{center}

\begin{center}
R.G.\ Roberts \\
Rutherford Appleton Laboratory, \\ Chilton, Didcot OX11 0QX, UK.
\\ \end{center}

\vspace*{0.75cm}

\begin{abstract}
We re-examine the estimates of the higher twist contributions to
the integral of $g_1$, the polarised structure function of the
nucleon, based on QCD sum rules. By including corrections both
to the perturbative contribution and to the low energy
contribution we find that the  matrix elements of the relevant
operators are more stable to variations of the Borel parameter
$M^2$, allowing for a meaningful estimate of the matrix elements.
We find that these matrix elements are typically twice as large
as previous estimates. However, inserting these new estimates
into the recently corrected expressions for the first moments of
$g_1$ leads to corrections too small to affect the
phenomenological analysis. For the unpolarised case the higher
twist corrections to the GLS and Bjorken sum rules are
substantial and bring the estimate of $\Lambda_{QCD}$ from the
former into good agreement with that obtained from the $Q^2$
dependence of deep inelastic data.
\end{abstract}

\newpage

The recent measurements of the polarised nucleon structure
functions $g_1^{p,n,d}$ \cite{EMC}, \cite{SLACE142}, \cite{NMC}
raise the question of the consistency of the three measurements
$I_{p,n,d} =\int^1_0 dx g_1^{p,n,d}(x)$.  In particular
since the three measurements are at different values of
$< Q^2 >$,
namely 10.7, 2 and 4.6 GeV$^2$ for $p,n$ and $d$, it is crucial
to understand the $Q^2$ dependence of the first moment of $g_1$
in order to test the Bjorken sum rule \cite{Bj} or to extract an
estimate of the nucleon's spin content, $\Delta q$, consistent
with all three experiments.

The higher order corrections to the leading twist expressions
have been calculated to $O(\alpha^2_s$) for the non-singlet
quantities and to $O(\alpha_s$  )
for the singlet contribution but it is the power corrections
$\frac{1}{Q^2}$ from
the higher twist operators which have recently \cite{EK},
\cite{CR} been shown may play an important role in a consistent
analysis of the data.  The magnitudes of the reduced
matrix elements of the relevant higher twist operators $U^S,
U^{NS}, V^S, V^{NS}$ were extracted from a QCD sum rule
calculation by Balitsky, Braun and Kolesnichenko (BBK) \cite{BBK}
and used in the analysis of ref \cite{EK}.
The aim of the present paper is to sharpen the results of BBK by
re-examining the computation of $\ll U^{S,NS}\gg$, $\ll
V^{S,NS}\gg$, including a contribution to   the
perturbative QCD side of the sum rule which was dropped in the
Borel transformation and explicitly retaining the continuum term
to the nucleon pole on the
low-energy side of the sum rule.  This leads to a significant
improvement in the stability of the extracted value of the
reduced matrix elements and hence to a more reliable estimate of
the higher twist contribution to the integral of $g_1$. Moreover
the estimated values of the matrix elements are significantly
larger than previously but since the coefficient of the twist
three piece in the first moment moment of $g_1$ has very recently
been corrected\cite{braun}, it turns out that the net higher
twist contribution to the moment is minimal.
The improvement in stability and increased magnitude of the
matrix elements is also true for the unpolarised case and we find
that the correction to the Gross-Llewellyn Smith (GLS) sum
rule\cite{GLS} is  sufficient to affect the extracted value of
$\Lambda_{\overline{MS}}$ substantially.

Following the procedure of BBK we consider the quantity
$\Gamma_\mu (p)$, if we are interested in the operators
$U^{S,NS}$, given by \begin{equation}
\Gamma_\mu (p) = i^2\int dx \; e^{ipx} \int dy \;
\langle T[\eta (x) U_\mu (y) \bar{\eta}(0)]\rangle
\label{eq:1}
\end{equation}
where $\eta$ is the nucleon current.  Expanding $\Gamma_\mu (p)
$ in powers of $\frac{1}{p^2}$ gives
\begin{equation}
\Gamma_\mu (p) =p_\mu {\rlap /p} \gamma_5 \left [ Ap^4 \ln^2
(\frac{\mu^2}{-p^2}) +B\ln (\frac{\mu^2}{-p^2}) + \tilde{C}
(\frac{1}{-p^2}) \ln(\frac{\mu^2}{-p^2})
+ C (\frac{1}{-p^2}) +D (\frac{1}{p^4}) \right ]
\label{eq:2}
\end{equation}
The coefficients $A,..,D$ may be read off from eq(8) of BBK
corresponding to the QCD evaluation of $\Gamma_{\mu}$, including
non-perturbative effects due to QCD condensates.   The next step
is to Borel transform the coefficient of $p_\mu {\rlap / p}
\gamma_5$ in eq(\ref{eq:2}) which gives
\begin{equation}
p_\mu {\rlap / p} \gamma_5 \left [
2AM^2\int^{s_0}_0 ds\; e^{-s/M^2} s^2 \ln
(\frac{\mu^2}{s}) + BM^4 \{1-e^{-s_0/M^2}\}
- \tilde{C} M^2\{ C_E
+\ln (\frac{\mu^2_{\overline{MS}}}{M^2}) \} +CM^2 +D \right ]
\label{eq:3}
\end{equation}
which differs from eq(11) of BBK since we have explicitly carried
out the $p^2  $
integration of the $\frac{1}{p^2} \ln (\frac{\mu^2}{p^2})$
contribution instead  of
neglecting it\footnote{Balitsky et al argue that with the natural
choice, $\mu^2_{\bar{MS}}\sim -p^2\sim 1 GeV^2$, this term
vanishes. However the Borel transform integrates over $p^2$ so
this is, at best, an approximation. Since the effect of the term
can be explicitly included we choose not to make this
approximation.}.

To complete the sum rule the quantity $\Gamma_\mu (p)$ must also
be determined in terms of the nucleon and continuum
contributions. Balitsky et al.\cite{BBK} use the form
\begin{equation}
\Gamma_\mu (p) = -p_\mu {\rlap / p} \gamma_5 \;\left [
\frac{2\lambda^2_p\ll U\gg}{(p^2-m^2)^2} + \frac{X}{(p^2-m^2)}
\right ]
\label{eq:4}
\end{equation}
where the first term is the pure nucleon pole contribution and
the single pole term is added to allow for the interference of
the pole term with a continuum contribution. The Borel transform
of the coefficient of $p_\mu {\rlap / p} \gamma_5$ in
eqn(\ref{eq:4}) is
\begin{equation}
\left [\frac{2\lambda^2_p}{M^2} e^{-m^2/M^2} \ll U\gg \;-\; e^{-
m^2/M^2} X \; \right  ]
\label{eq:5}
\end{equation}
and the QCD sum rule results from equating eq(\ref{eq:3}) to
eq(\ref{eq:5}) to give
\begin{eqnarray}
\ll U \gg \; -\; XM^2 \;& =  & \left [
\frac{e^{m^2/M^2}}{2\lambda^2_p}\right ] \times \hfill  \left [
2AM^2\int^\infty_{s_0} ds\; e^{-s/M^2} s^2  \ln
(\frac{\mu^2}{s}) \right.
\nonumber \\
& & +  BM^4 \{1-e^{-s/M^2}\} \left .
- \tilde{C} M^2\{ C_E +\ln  (\frac{\mu^2_{\overline{MS}}}{M^2})
\} +CM^2 + D \right ]
\label{eq:6}
\end{eqnarray}

In this paper we are particularly concerned to estimate of the
errors in determining the operator matrix elements from the QCD
sum rules.  Thus we will consider in detail the effects of each
of the terms in this expansion and the inclusion of further terms
in the Borel expansion in $(M^2)^n/n!$.
On the rhs of eq(\ref{eq:2}) the first term not included is $\sim
1/{p^6}$ which, after Borel transformation contributes a term
$\sim 1/{2!M^2}$ to the rhs of eq(\ref{eq:6}).  The coefficient
of this term should be no bigger than the coefficient of Y if the
perturbative expansion in $1/(n!(M)^n)$ is acceptably convergent.
By adding such a term and fitting its coefficient we may check
this over some range of $M^2$ and hence establish over what range
of $M^2$ (if any) the sum rule converges.

The corrections to the lhs of eq(\ref{eq:6}) are somewhat more
difficult to determine.  Further resonance contributions with
mass ${m_R^2} > {m^2}$ can be included by adding a resonance pole
$(m_R^2 - p^2)^{-1}$.  After Borel transformation this gives an
additional term $M^2 e^{(m^2-m_R^2)/M^2}\; Y$ on the lhs of
eq(\ref{eq:6}).  However the dominant correction to the lhs of
eq(\ref{eq:2}) is not expected to come from a nucleon resonance
excitation but from the $(N + n \pi)$ continuum which has a
threshold at $E = \sqrt{m^2 + m_{\pi}^2}$, very close to the
nucleon pole.  As far as we know this contribution has not been
considered explicitly even though it is potentially very large.
However, we will demonstrate that this contribution does not
substantially degrade the accuracy with which we can
determine the operator matrix elements provided the resonance
term discussed above is added to eq(\ref{eq:6}).  The reason is
that the $\pi N$ contribution is well described by the terms
proportional to X  in eq(\ref{eq:4}) together with the resonance
term proportional to Y.
To demonstrate this we will estimate the contribution due to the
$N\pi$ intermediate state, fig.2.  This gives the term
\begin{equation}
I_{\pi N}=\frac{g_{\pi N N}^2
\lambda_p^2}{m^2}\int\frac{f(q^2)^2[-
2(p+q)_{\mu}(\rlap /p + \rlap /q)\gamma_5\ll U \gg]d^4q}
{(2\pi)^4(q^2-m_{\pi}^2)(m^2-(p+q)^2)^2}
\label{eq:7}
\end{equation}

Here we have included a form factor $f(q^2)$ needed to describe
the $\eta N \pi$ coupling far from the pion mass shell (on shell
we take $\langle \eta\mid N \pi \rangle = \lambda_p g_{\pi N
N}$).  The result is insensitive to the particular choice of
$f(q^2)$ provided it provides convergence for large $q^2$; here
we choose
$f(q^2) =q^2/((q^2-m_{\pi}^2)(1- q^2 /0.7 GeV^2))$ i.e. chosen
to vanish as $q^2\rightarrow 0$ and to be the same as the nucleon
electromagnetic form factor for large spacelike $q^2$. (Since
$f(q^2)$ is needed for spacelike $q^2$ we have chosen a form
factor with no singularities in the spacelike region.) After
taking the Borel transform of the coefficient of $p_\mu \rlap /p
\gamma_5$ and  multiplying by $M^2e^{m^2/M^2}/2\lambda^2_p$
we find the resulting contribution to the lhs of eq(\ref{eq:6})
is accurately reproduced by the form
\begin{equation}
\alpha+\beta M^2+ \gamma M^2 e^{(m^2-m_R^2)/M^2}
\label{eq:8}
\end{equation}
with $\alpha \sim 0.03 \ll U \gg$,
$\beta \sim 0.15 \ll U \gg$,
$\gamma \sim 0.2 \ll U \gg$,
and $m_R^2\approx 2m^2$.  The
origin of this form is easy to understand. The $\pi N$
singularity starts at $(m_{\pi}+m)$ in the energy plane and,
after Borel transformation, gives a contribution proportional to
the integral of $e^{(m^2-\tilde{m}^2)}$ over $\tilde{m}$, with
an appropriate weighting factor. To a good approximation this may
be approximated by the sum of two exponentials $e^{(m^2-
m_1^2)/M^2}$ and $e^{(m^2-m_R^2)/M^2}$ with
$m_1\approx(m_{\pi}+m)$. Taylor expanding the first term in
$m_{\pi}$ leads to eq(\ref{eq:8}). Thus we see that the $\pi N$
contribution may, to a good
approximation, be included in the continuum and interference
terms and gives a relatively small correction to the
determination of the operator matrix elements.  Using similar
methods  suggests that the contribution to constant term in
eq(\ref{eq:8}) from $N+n \pi$ intermediate states will also be
quite small.

  Let us now turn to the phenomenological analysis and first
consider the analysis of ref\cite{BBK}.
We have four operators to consider $U^S, U^{NS}, V^S, V^{NS}$ the
latter two having $p_\mu {\rlap / p} \gamma_5$ replaced by
$S_{\nu,\sigma} A_{\mu,\nu} p_\sigma p_\mu \gamma_\nu \gamma_5$
where $A,S$ stand for symmetric and antisymmetric combination of
indices.  So we have four QCD sum rules with coefficients
$A\rightarrow E$ given by BBK eqs(8,9).
In extracting a value of  $\ll U\gg$ from the sum rule, BBK
retained only the first two
terms on the lhs and also set $\tilde{C}$=0. They obtained  $\ll
U\gg$ at each value of $M^2$ by applying
$(1-M^2\frac{d}{dM^2})$ to the rhs.  The stability of this
estimate of $\ll U\gg$ relies
upon the rhs of eq(\ref{eq:6}) being approximately linear over
a range of $M^2$.
The solid lines in fig.1 are the BBK values for the matrix
elements extracted by this procedure  and the variation with
$M^2$ reflects the importance of higher derivatives in $M^2$,
making it difficult to arrive at a reliable estimates for $\ll
U,V \gg$.

Let us see how the situation changes with the addition of the
proposed continuum term  $M^2 e^{(m^2-m_R^2)/M^2}\; Y$
and the term proposal to
$\tilde{C}$ in eq(\ref{eq:6}).  As we will demonstrate these are necessary
to describe adequately the $M^2$ behaviour of the rhs over a
reasonable range.
We first fit the rhs of eq(\ref{eq:6}) by the form on the lhs
with four parameters $\ll U \gg$, $X$, $Y$ and $m_R^2$ (the
last  corresponding to an effective threshold for the continuum)
over a wide range  0.5 GeV$^2 < M^2 < 2.0$ GeV$^2$.
Then, with  $m_R^2$ fixed (typically around 2 GeV$^2$),
we perform a three parameter fit over each
interval of 0.2 GeV$^2$ in $M^2$. The dashed line in fig.2 shows
the  resulting values of $\ll U \gg$ and we immediately see a
marked improvement in the stability compared to neglecting the
continuum term which results in a more meaningful
estimate of the matrix elements. At the same time we notice that
the actual value obtained is considerably different. At $M^2 =
1$ GeV$^2$ for example the new estimate is roughly twice the old
value. If instead we choose to describe the continuum by
$e^{(m^2-m_R^2)/M^2}\; Y$ (i.e. a double resonance pole) there is
a similar improvement in stability and the extracted values of
the matrix elements are within 15\% of those in fig.1.

Using these results we will now try to sharpen the estimates of
the higher twist contributions further by considering possible
additional terms to the rhs of eq(\ref{eq:6}).  We will
consider the sensitivity of $\ll U, V\gg$ to an additional term
$\sim \frac{1}{p^6}$ on the rhs of eq(\ref{eq:2}), i.e. a term
$\sim \frac{1}{2M^4}$ on the rhs of eq(\ref{eq:6}).
Thus we may have
an additional term $\frac{e^{m^2/M^2}}{2M^2}$
on the lhs of eqn (6) which can be used in fitting
the rhs and help determine the
range of $M^2$ over which the sum rule is valid.  As a result of
fitting the curves (with $\mu^2=$ 1 GeV$^2$) in fig.1 with the
constraint that the coefficient of the additional term is no
bigger than the coefficient $Y$ of the continuum term
lead to an acceptable range 0.8 $\leq M^2 \leq$ 1.75 GeV$^2$.

In fact, the magnitude of the continuum is
correlated, as one might expect, with the value of the hadronic
continuum parameter $s_0$ on the rhs. Our results
correspond to $s_0 = 2.25$ GeV$^2$ as in refs\cite{BBK,BK,Iof}.
The dependence on $s_0$ is weak however and careful fitting
reveals that this dependence is absorbed by the explicit
continuum contributions, leaving the magnitude of $\ll U, V \gg$
practically invariant as $s_0$ varies in the range 1.8 to 5
GeV$^2$.

For the above range in $M^2$, the resulting uncertainties in the
values of $\ll  U, V\gg$
from the fitting procedure are comparable to the uncertainties
expected from varying $\mu^2$ in the range 0.33 to 3 GeV$^2$.
The values of the reduced matrix elements obtained are
\begin{eqnarray}
\ll U^S\gg  = 0.046 \pm  0.010 GeV^2 \quad\quad \ll U^{NS}\gg =
0.317 \pm 0.010 GeV^2\nonumber\\
m^2 \ll V^S\gg  = -0.292 \pm  0.010 GeV^2\;\;\; m^2 \ll V^{NS}\gg
= 0.605 \pm 0.030 GeV^2
\label{eq:9}
\end{eqnarray}

{}From the values in eq(\ref{eq:9}) we compute the coefficients
$a_p$ and $a_n$ of the $1/Q^2$ contributions to the integrals of
$g^p_1$ and $g^n_1$, using the corrected formulas of
ref\cite{braun} \begin{eqnarray}
a_p + a_n &=& -\frac{8}{9} . \frac{5}{18} \; \left [\ll U^S\gg -
 \frac{1}{4} m^2 \ll V^S\gg \right ]\nonumber\\
a_p - a_n &=& -\frac{8}{9} . \frac{1}{6} \; \left[\ll U^{NS}\gg -
 \frac{1}{4} m^2 \ll V^{NS}\gg \right ]
\label{eq:10}
\end{eqnarray}
which gives
\begin{equation}
a_p = - 0.029 \pm 0.002,\;\;\; a_n = - 0.002 \pm 0.002.
\label{eq:11}
\end{equation}
The errors in eq(\ref{eq:9})are
based simply on the small variation of the values of
$\ll U,V \gg$ with $M^2$ and $\mu ^2$ and are typically 5\%.  These
errors are in addition to the underlying uncertainties arising from
the factorisation assumption for the vacuum condensates which are
typically 20\%\cite{vlbpc}. Thus a realistic estimate of the
errors in eq(\ref{eq:11})
is more like 0.010.

The integrals $I_{p,n,d}$ can be written
\begin{eqnarray}
I_p &=& I_3 + I_8 + I_0 + a_p/Q^2\nonumber\\
I_n &=& - I_3 + I_8 + I_0 + a_n/Q^2\nonumber\\
I_d &=&  I_8 + I_0 + (a_p+a_n)/Q^2
\label{eq:12}
\end{eqnarray}
where
\begin{eqnarray}
I_3 &=& \;\frac{1}{12} \;[F+D]\; [1-\frac{\alpha_s}{\pi} - 3.58
(\frac{\alpha_s}{\pi})^2]\nonumber\\
I_8 &=& \;\frac{1}{36} \;[3F-D]\; [1-\frac{\alpha_s}{\pi} - 3.58
(\frac{\alpha_s}{\pi})^2]\nonumber\\
I_0 &=& \frac{1}{9}\;\Delta q \;   [1- \frac{\alpha_s}{3\pi}]
\label{eq:13}
\end{eqnarray}

Using the measured values of the polarisation asymmetries from
refs \cite{EMC,SLACE142,NMC}, the values of $I_{p,n,d}$ at values
of $Q^2$ = 10.7, 2, 4.6 GeV$^2$ were extracted in ref \cite{CR}
and determined to be 0.134$\pm$0.012, $-$0.023$\pm$0.005,
0.041$\pm$0.016 respectively. The estimates for the coefficients
$a_p,
a_n$ from our improved QCD sum rule analysis, eq(\ref{eq:11}),
when inserted into eqs(\ref{eq:12},\ref{eq:13}) yield estimates
for the nucleon spin content (for $F/D$ = 0.575 $\pm$ 0.016)
\begin{eqnarray}
\Delta q & =& 0.24 \pm 0.11 \;\;\; {\rm from} \; p\nonumber\\ &
=& 0.53 \pm 0.07 \;\;\; {\rm from} \; n\nonumber\\
& =& 0.27 \pm 0.15 \;\;\; {\rm from} \; d
\label{eq:14}
\end{eqnarray}

The value of $\Delta q$ obtained from the relatively low $Q^2$
neutron data from SLAC is still out of line with the values
obtained form CERN on the proton and deuterium. If the neutron
higher twist coefficient $a_n$ had come out large and positive,
around 0.04 or so, then $\Delta q$ would decrease a value close
to the proton and deuteron estimates. In fact the analysis of
Ellis and Karliner\cite{EK} used such a value based on the
BBK\cite{BBK} estimates of $\ll U,V \gg$ but with the (now known
to be) incorrect formulas for twist three contribution to the
first moment. Thus despite the fact that we claim considerably
larger estimates for the matrix elements, the corrected
formulas\cite{braun} lead to a small neutron correction, thereby
ruling out higher twists as a way of reconciling the three
experiments. Bag model estimates for $a_n$ give a zero
value\cite{Ji}.

The improvement to our understanding of the QCD sum rule
estimates of $\ll U, V\gg$ has led to a more meaningful
determination of the higher twist corrections to the integrals
of the polarised structure function $g_1$.
We recall that two
corrections to the sum rule $-$ the evaluation of the
$\frac{1}{p^2} \ln (\frac{\mu^2}{p^2})$ contribution on the rhs
and the inclusion of the continuum contribution $-$ resulted in
more stable estimates of the reduced matrix elements.  It is
therefore natural to ask if similar corrections apply to other
sum rules e.g. the matrix elements $\ll O^{S,NS}\gg$ which
determine the $1/Q^2$ corrections to the GLS\cite{GLS}
and Bjorken ($F_1$)\cite{Bjf1}
sum rules \cite{BK}.  The first correction does not apply since
$\frac{1}{p^2} \ln (\frac{\mu^2}{p^2})$ terms cancel for the
unpolarised operator
$p_\mu {\rlap / p}$ but the continuum correction should be
included.

Fig.3 shows the corresponding determinations of $\ll
O^{S,NS}\gg$. In fig.3(a) we note the strong variation of $\ll
O^S \gg$ with $M^2$, again indicating the inadequacy of the
nucleon pole terms alone in eq(\ref{eq:4}). Carrying out an
analogous fitting procedure as in the polarised case, the
resulting estimate for $\ll O^S \gg$ is far less sensitive to the
value of $M^2$ indicated by the dashed line in fig.3(a).
Interestingly, the estimated
magnitude of both matrix elements
$\ll O^{S,NS} \gg$ increases when a realistic continuum term is
included. In particular, the estimate used in the analysis of
Chyla and Kataev\cite{CK} was that of ref\cite{BK}
\begin{equation}
\ll O^S \gg = 0.33\pm 0.16 GeV^2
\label{eq:15}
\end{equation}
which led to a value of $\Lambda_{\overline{MS}}$ extracted from
the data on $xF_3(x,Q^2=3$ GeV$^2$ on the GLS sum rule of
\begin{equation}
\Lambda_{\overline{MS}}^{(4)} =
318\pm23(stat)\pm99(syst)\pm62(twist) MeV.
\label{eq:16}
\end{equation}
If $I_{GLS} $ is the measured value of the GLS sum rule, then
\begin{equation}
1 - \frac{I_{GLS}}{3} = \frac{\alpha_s(Q^2)}{\pi} + \frac{8}{27}
\frac{\ll O^S \gg}{Q^2}
\label{17}
\end{equation}
and we see that an  increased estimate of $\ll O^S \gg$ leads to
a lower value of $\Lambda_{\overline{MS}}$. We estimate the
larger and more precise value
\begin{equation}
\ll O^S \gg = 0.53\pm 0.04 GeV^2
\label{eq:18}
\end{equation}
which leads to
\begin{equation}
\Lambda_{\overline{MS}}^{(4)} =
232\pm23(stat)\pm99(syst)\pm17(twist) MeV
\label{eq:19}
\end{equation}
which is more in accord with estimates got from studying the
$Q^2$ dependence of deep inelastic data\cite{pdu},
$\Lambda_{\overline{MS}}^{(4)} = 230\pm55$ MeV. Again, the error
in eq(\ref{eq:18}) does not include the intrinsic uncertainty ($\sim$
20\%) associated with the factorisation assumption; including this
raises the final error in eq(\ref{eq:19}) from 17 to 45 MeV.

In summary, we have shown that there are corrections to QCD sum
rules which have not been included in previous determinations of
the relevant reduced matrix elements. These corrections lead to
significant improvement in the stability of the extracted values
with respect to the range of the Borel parameter $M^2$. As an
example, we have applied these corrections to the case of the
polarised structure function $g_1$ of the nucleon and found that
the size of the the higher twist contributions is now determined
with better precision. We have considered in some detail the
corrections to the sum rule. The $N\pi$ correction, potentially
very significant, has been shown to have little effect on the
determination of the reduced matrix elements. The remaining
contributions are under control for a reasonable range of $M^2$
allowing for a realistic determination of the error.
Our final estimates for the matrix elements, both for polarised
and unpolarised structure functions are more reliable and,
moreover, significantly larger than previous estimates.
Nevertheless because of the recent corrections to the term
multiplying the twist three contribution, the resulting impact
on the phenomenology of moments of $g_1^{p,n,d}$ is reduced. For
the analysis of the GLS sum rule  however, the resulting value
of $\Lambda_{\overline{MS}}^{(4)}$ is more in line with other
deep inelastic scattering phenomenology.

\vspace*{1cm}
\noindent {\large \bf Acknowledgements}
\vspace*{.5cm}

We are grateful for correspondence with Vladimir Braun.

\vskip 1cm
%\newpage
\noindent{\large\bf Figure Captions}

\begin{itemize}
\item[Fig.\ 1] Values (in GeV$^2$) of the matrix elements  $\ll
U,V \gg$ derived from
fitting the rhs of eq(\ref{eq:6}) in bins of 0.2 GeV$^2$ in $M^2$
with just the two terms on the lhs (solid lines) and together with
the continuum term $M^2 e^{(m^2-m_R^2)/M^2}\; Y$(dashed lines).

\item[Fig.\ 2] $\pi$ N contribution, eq(\ref{eq:7}), to the
correlation function.

\item[Fig.\ 3] Values (in GeV$^2$) of the matrix elements $\ll
O^{S,NS} \gg$ derived from fitting the rhs of eq(39) of
BK\cite{BK} with the nucleon double and single pole terms only
(solid lines) and together with a  continuum term (dashed lines),
as in fig.1.

\end{itemize}

\end{document}